\documentstyle[aps,epsf,epsfig,prl,psfig]{revtex}
\newcommand{\beq}{\begin{equation}}
\newcommand{\eeq}{\end{equation}}
\newcommand{\beqas}{\begin{eqnarray*}}
\newcommand{\eeqas}{\end{eqnarray*}}
\newcommand{\beqar}{\begin{eqnarray}}
\newcommand{\eeqar}{\end{eqnarray}}

\newcommand{\subs}[1]{{\mbox{\scriptsize #1}}}

\begin{document}
\newcommand \be {\begin{equation}}
\newcommand \ee {\end{equation}}

\twocolumn[\hsize\textwidth\columnwidth\hsize\csname
@twocolumnfalse\endcsname
%\title{Random Matrix Theory to the rescue?}
\title{Noise Dressing of Financial Correlation Matrices}
\author{Laurent Laloux$^\dagger$, Pierre Cizeau$^\dagger$, Jean-Philippe
Bouchaud$^{\dagger,*}$ and Marc Potters$^\dagger$}
\address{$^\dagger$ Science \& Finance, 109-111 rue Victor Hugo, 92532
Levallois Cedex, FRANCE}
\address{$^*$ Service de Physique de l'\'Etat Condens\'e,
 Centre d'\'etudes de Saclay, \\ Orme des Merisiers, 
91191 Gif-sur-Yvette C\'edex, FRANCE\\}
\date{\today}
\maketitle
\widetext

\begin{abstract}
We show that results from the theory of random matrices are
potentially of great interest to understand the statistical structure of
the empirical correlation matrices appearing in the study of price
fluctuations. The central result of the present study is the
remarkable agreement between the theoretical prediction (based on the
assumption that the correlation matrix is random) and empirical
data concerning the density of eigenvalues associated to the time
series of the different stocks of the S\&P500 (or other major
markets). In particular the present study raises serious doubts
on the blind use of empirical correlation matrices for risk management.
\end{abstract}

\narrowtext

]
An important aspect of risk management is the estimation of the
correlations between the price movements of different assets. The
probability of large losses for a certain portfolio or option book is
dominated by correlated moves of its different constituents -- for
example, a position which is simultaneously long in stocks and short
in bonds will be risky because stocks and bonds move in opposite
directions in crisis periods. The study of correlation (or covariance)
matrices thus has a long history in finance, and is one of the
cornerstone of Markowitz's theory of optimal portfolios \cite{CAPM}.
However, a reliable empirical determination of a correlation matrix
turns out to be difficult: if one considers $N$ assets, the
correlation matrix contains $N(N-1)/2$ entries, which must be
determined from $N$ time series of length $T$; if $T$ is not very
large compared to $N$, one should expect that the determination of the
covariances is noisy, and therefore that the empirical correlation
matrix is to a large extent random, i.e.\ the structure of the matrix
is dominated by measurement noise. If this is the case, one should be
very careful when using this correlation matrix in applications.  In
particular, as we shall show below, the smallest eigenvalues of this
matrix are the most sensitive to this `noise' -- on the other hand, it
is precisely the eigenvectors corresponding to these smallest
eigenvalues which determine, in Markowitz theory, the least risky
portfolios \cite{CAPM}. It is thus important to devise methods which
allows one to distinguish `signal' from `noise', i.e.\  eigenvectors
and eigenvalues of the correlation matrix containing real information
(which one would like to include for risk control), from those which
are devoid of any useful information, and, as such, unstable in time.
From this point of view, it is interesting to compare the properties
of an empirical correlation matrix {\bf C} to a `null hypothesis'
purely {\it random}\/ matrix as one could obtain from a finite time
series of strictly uncorrelated assets.
Deviations from the random matrix case might then suggest the presence 
of true information. The theory of Random Matrices has a long history 
in physics since the fifties \cite{WigDys}, and many results are known 
\cite{Mehta}. As shown below, these results are also of
genuine interest in a financial context (see also \cite{Galluccio}). 

The empirical correlation matrix {\bf C} is constructed from the time 
series of price changes\footnote{In the following we assume that the average
value of the $\delta x$'s has been subtracted off, and that the $\delta x$'s 
are rescaled to have a constant unit volatility.} $\delta x_i(t)$
(where $i$ labels the asset and $t$ the time) through the equation:
\be\label{defcor}
\mbox{{\bf C}}_{ij} = \frac{1}{T}\sum_{t=1}^T \delta x_i(t) \delta x_j(t).
\ee

We can symbolically write Eq.\ (\ref{defcor}) as 
{\bf C}$=$ 1/T {\bf M\, M}$^{\mbox{\scriptsize T}}$,
where {\bf M} is a $N \times T$ rectangular matrix, and
$^{\mbox{\scriptsize T}}$ denotes matrix transposition.  The null hypothesis 
of uncorrelated assets, which we consider now, translates itself in 
the assumption that the coefficients $M_{it}=\delta x_i(t)$ are independent, 
identically distributed, random variables\footnote{Note
that even if the `true' correlation matrix 
{\bf C}$_\subs{true}$ is the identity matrix, its empirical determination 
from a finite time series will generate non trivial 
eigenvectors and eigenvalues.}. 
We will note $\rho_C(\lambda)$ the density of eigenvalues of {\bf C}, 
defined as:
\be
\rho_C(\lambda)=\frac{1}{N} \frac{d n(\lambda)}{d \lambda},
\ee
where $n(\lambda)$ is the number of eigenvalues of {\bf C} less than $\lambda$. 
Interestingly, if {\bf M} is a $T \times N$ random matrix, 
$\rho_C(\lambda)$ is exactly known in the limit
$N \to \infty$, $T \to \infty$ and $Q=T/N\geq1$ fixed \cite{Sengupta}, and reads:
\beqar\nonumber
\rho_C(\lambda)&=&\frac{Q}{2\pi\sigma^2}
\frac{\sqrt{(\lambda_\subs{max}-\lambda)(\lambda-\lambda_\subs{min})}}{\lambda},\\
\lambda_\subs{min}^\subs{max}&=&\sigma^2(1+1/Q\pm 2\sqrt{1/Q}),\label{rho}
\eeqar
with $\lambda \in 
[\lambda_\subs{min},\lambda_\subs{max}]$, 
and where $\sigma^2$ is equal to the variance of 
the elements of {\bf M} \cite{Sengupta}, equal to $1$ with our normalisation.
In the limit $Q=1$  the normalised eigenvalue
density of the matrix ${\bf M}$ is the well known Wigner semi-circle law,
and the corresponding distribution of the {\it square} of these eigenvalues 
(that is, the eigenvalues of {\bf C}) is then indeed given by (\ref{rho}) 
for $Q=1$.  The most important features predicted by Eq.\ (\ref{rho}) are:
\begin{itemize}
\item the fact that the lower `edge' of the spectrum is strictly positive 
(except for $Q=1$); there is therefore no eigenvalues between 
$0$ and $\lambda_\subs{min}$.  Near this edge, the density of 
eigenvalues exhibits a sharp maximum, except in
the limit $Q=1$ ($\lambda_\subs{min}=0$) where it diverges as 
$\sim {1}/{\sqrt{\lambda}}$.

\item the density of eigenvalues also vanishes above a certain upper edge
$\lambda_\subs{max}$.
\end{itemize}

Note that the above results are only valid in the limit $N \to \infty$.
For finite $N$, the singularities present at both edges are smoothed: 
the edges become somewhat blurred, with a small
probability of finding eigenvalues above $\lambda_\subs{max}$ 
and below $\lambda_\subs{min}$, which goes to zero when $N$ becomes large.

Now, we want to compare the empirical distribution of the eigenvalues of the 
correlation matrix of stocks corresponding to different markets with the
theoretical prediction given by Eq.\ (\ref{rho}), 
based on the assumption that the correlation matrix is random.  
We have studied numerically the density of eigenvalues of the correlation 
matrix of $N=406$ assets of the S\&P 500, 
based on daily variations during the years 1991-96, 
for a total of $T=1309$ days (the corresponding value of $Q$ is $3.22$). 

An immediate observation is that the highest eigenvalue $\lambda_1$ is $25$
times larger than the predicted $\lambda_\subs{max}$ -- see Fig.\ 1, inset. 
(The corresponding eigenvector is, as expected,  the `market' itself, 
i.e. it has roughly equal components on all the $N$ stocks.) 
The simplest `pure noise' hypothesis is therefore inconsistent with the
value of $\lambda_1$. A more reasonable idea is that the components of 
the correlation matrix which are orthogonal to the `market' is pure noise. 
This amounts to subtracting the contribution of $\lambda_\subs{max}$ 
from the nominal value $\sigma^2=1$, leading to 
$\sigma^2=1-\lambda_\subs{max}/N=0.85$. 
The corresponding fit of the empirical distribution is shown as 
a dotted line in Fig.\ 1.
Several eigenvalues are still above $\lambda_\subs{max}$ and might contain 
some information, thereby reducing the variance of the effectively 
random part of the correlation matrix. One can therefore treat $\sigma^2$ as an
adjustable parameter. The best fit is obtained for $\sigma^2=0.74$, 
and corresponds to the dark line in Fig.\ 1, which accounts quite 
satisfactorily for 94\% of the spectrum, while the 6\% highest 
eigenvalues still exceed the theoretical upper edge by a substantial amount.
Note that still a better fit could be obtained by allowing for a 
slightly smaller effective value of $Q$, which could account for 
the existence of volatility correlations \cite{ustocome}. 

We have repeated the above analysis on different stock markets (e.g. Paris)
and found very similar results. 
In a first approximation, the location of the theoretical edge, determined by
fitting the part of the density which contains most of the eigenvalues, allows
one to distinguish `information' from `noise'. 
However, a more careful study should be
undertaken, in particular to treat adequately the finite $N$ effects. 
\vskip 0.2cm
\begin{figure}
\centerline{\psfig{figure=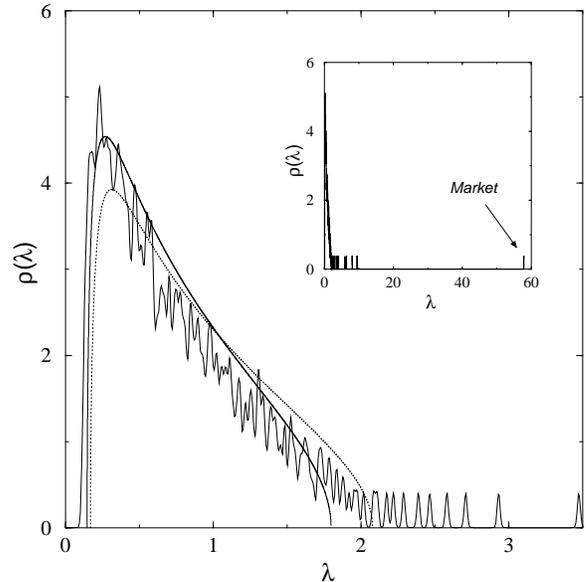,width=8cm}}
\vskip 0.8cm
\caption{Smoothed density of the eigenvalues of {\bf C}, where
the correlation matrix {\bf C} is extracted from $N=406$ assets of the
S\&P500 during the years 1991-1996.  For comparison we have plotted
the density Eq.\ (6) for $Q=3.22$ and $\sigma^2=0.85$: this is the theoretical
value obtained assuming that the matrix is purely random except for
its highest eigenvalue (dotted line). A better fit can be obtained
with a smaller value of $\sigma^2=0.74$ (solid line), corresponding to 
74\% of the total variance.  Inset: same plot,
but including the highest eigenvalue corresponding to the `market',
which is found to be 30 times greater than
$\lambda_\subs{max}$.} \label{fig1}
\end{figure}

The idea that the low lying eigenvalues are essentially random can also be
tested by studying the statistical structure of the corresponding 
{\it eigenvectors}.  The $i^{th}$ component
of the eigenvector corresponding to the eigenvalue $\lambda_\alpha$ will be
denoted as $v_{\alpha,i}$. We can normalise it such that $\sum_{i=1}^N
v_{\alpha,i}^2 =N$. If there is no information contained in the eigenvector
$v_{\alpha,i}$, one expects that for a fixed $\alpha$, the distribution of
$u=v_{\alpha,i}$ (as $i$ is varied) is a maximum entropy distribution, such
that $\overline{u^2}=1$. This leads to the so-called Porter-Thomas
distribution in the theory of random matrices:
\be
P(u) = \frac{1}{\sqrt{2\pi}} \exp -\frac{u^2}{2}\label{PT}.
\ee
As shown in Fig.\ 2, this distribution fits extremely well the empirical
histogram of the eigenvector components, except for those corresponding 
to the highest eigenvalues, which lie beyond the theoretical edge 
$\lambda_\subs{max}$.  We show in the inset the distribution of $u$'s 
for the highest eigenvalue, which is markedly different from 
the `no information' assumption, Eq.\ (\ref{PT}). 

We have finally studied correlation matrices corresponding not to
price variations but to the (time dependent) volatilities of the
different stocks, determined from the study of intraday
fluctuations. These matrices should contain some relevant information
for option trading and hedging. The obtained results are again very
similar to those shown in Fig.\ 1 and 2.

\vskip 0.2cm
\begin{figure}
\centerline{\psfig{figure=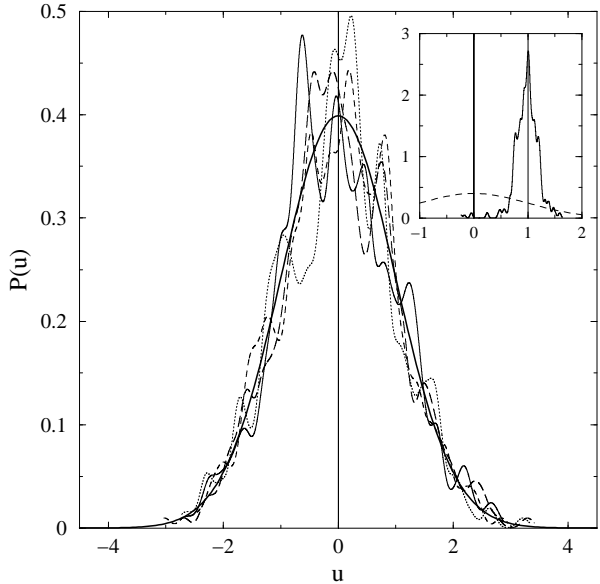,width=8cm}}
\vskip 0.8cm
\caption{
Distribution of the eigenvector components $u=v_{\alpha,i}$, for five different
eigenvectors well inside the interval $[\lambda_\subs{min},\lambda_\subs{max}]$, 
and comparison with the `no information' assumption, 
Eq.\ (\protect\ref{PT}). Note that there are {\it no}\/ adjustable parameters.
Inset: Plot of the same quantity for the highest eigenvalue, 
showing marked differences with
the theoretical prediction (dashed line), which is indeed expected. \label{fig2}}
\end{figure}

To summarise, we have shown that results from the theory of random
matrices (well documented in the physics literature \cite{Mehta}) is
of great interest to understand the statistical structure of the
empirical correlation matrices. The central result of the present
study is the remarkable agreement between the theoretical prediction
and empirical data concerning both the density of eigenvalues and the
structure of eigenvectors of the empirical correlation matrices
corresponding to several major stock markets.  
Indeed, in the case of the S\&P 500,  94\% of the
total number of eigenvalues fall in the region where the theoretical
formula (\ref{rho}) applies. Hence, less than 6\% of the eigenvectors 
which are responsible of 26\% of the total volatility,
appear to carry some information. This method might be very useful to
extract the relevant correlations between financial assets of various
types, with interesting potential applications to risk management and
portfolio optimisation. It is clear from the present study that
Markowitz's portfolio optimisation scheme based on a purely historical
determination of the correlation matrix is not adequate, since its
lowest eigenvalues (corresponding to the smallest risk portfolios) are
dominated by noise.

Acknowledgements: We want to thank J.P. Aguilar (CFM), J.M. Lasry
(Paribas), M. Meyer (Science \& Finance) and S. Galluccio (Paribas)
for discussions.


\begin{thebibliography}{99}

\bibitem{CAPM} E.J. Elton and M.J. Gruber, {\it Modern Portfolio Theory and
Investment Analysis} (J.Wiley and Sons, New York, 1995);
H. Markowitz, {\it Portfolio Selection: Efficient Diversification
of Investments} (J.Wiley and Sons, New York, 1959). See also: J.P. Bouchaud
and M. Potters, 
{\it Theory of Financial Risk}, (Al\'ea-Saclay, Eyrolles, Paris, 1997) (in
french). 
\bibitem{WigDys} For a review, see: O. Bohigas, M. J. Giannoni, {\it
Mathematical and computational methods in nuclear physics}, Lecture Notes in
Physics, Vol. 209, Springer-Verlag (1983)
\bibitem{Mehta} M. Mehta, {\it Random Matrices} (Academic Press, New York,
1995).
\bibitem{Galluccio} S. Galluccio, J.P. Bouchaud and M. Potters, Physica {\bf A 259}, 449 (1998).
\bibitem{Sengupta} A.M. Sengupta and P.P. Mitra
{\it Distribution of Singular Values for Some Random Matrices},
cond-mat/9709283 preprint.
\bibitem{ustocome} L. Laloux, P. Cizeau, J.P. Bouchaud and M. Potters, in preparation.

\end{thebibliography}
\end{document}